\documentclass[conference]{IEEEtran}
\IEEEoverridecommandlockouts
\usepackage{cite}
\usepackage{amsmath,amssymb,amsfonts}
\usepackage{algorithmic}
\usepackage{graphicx}
\usepackage[table,xcdraw]{xcolor}
\usepackage{textcomp}
\usepackage{xcolor}
\usepackage{flushend}
\def\BibTeX{{\rm B\kern-.05em{\sc i\kern-.025em b}\kern-.08em
    T\kern-.1667em\lower.7ex\hbox{E}\kern-.125emX}}
\begin{document}

\title{Analysis of EEG frequency bands for Envisioned Speech Recognition}

\author{\IEEEauthorblockN{Ayush Tripathi}
\IEEEauthorblockA{\textit{Department of Electrical Engineering} \\
\textit{Indian Institute of Technology Delhi}\\
New Delhi, India \\
eez208477@ee.iitd.ac.in}
}

\maketitle

\begin{abstract}

The use of Automatic speech recognition (ASR) interfaces have become increasingly popular in daily life for use in interaction and control of electronic devices. The interfaces currently being used are not feasible for a variety of users such as those suffering from a speech disorder, locked-in syndrome, paralysis or people with utmost privacy requirements. In such cases, an interface that can identify envisioned speech using electroencephalogram (EEG) signals can be of great benefit. Various works targeting this problem have been done in the past. However, there has been limited work in identifying the frequency bands ($\delta, \theta, \alpha, \beta, \gamma$) of the EEG signal that contribute towards envisioned speech recognition. Therefore, in this work, we aim to analyze the significance of different EEG frequency bands and signals obtained from different lobes of the brain and their contribution towards recognizing envisioned speech. Signals obtained from different lobes and bandpass filtered for different frequency bands are fed to a spatio-temporal deep learning architecture with Convolutional Neural Network (CNN) and Long Short-Term Memory (LSTM). The performance is evaluated on a publicly available dataset comprising of three classification tasks - digit, character and images. We obtain a classification accuracy of $85.93\%$, $87.27\%$ and $87.51\%$ for the three tasks respectively. The code for the implementation has been made available at https://github.com/ayushayt/ImaginedSpeechRecognition.

\end{abstract}


\maketitle

\section{Introduction}

The emergence of ubiquitous electronic devices has increased the requirement for natural ways of interaction with electronic devices. This has in turn led to significant improvement in human computer interfaces (HCI) such as those driven by speech and gesture \cite{herff2016automatic}. However, there are certain situations in which the use of such mediums of HCI are limited. This may be due to speech disabilities, conditions such as locked-in syndrome or situations in which privacy is of utmost priority to the user. In such cases, imagined speech recognition systems based on non-invasive Electroencephalography could be used as an exciting alternative. Through years, various Brain Computer Interfaces (BCIs) such as those on the principle of text entry or imagination of motor movements have been proposed. Most of these however are focused on a user who is required to focus on a particular image or letter for a significant duration of time in order to select it or imagine a particular movement for providing commands \cite{sreeja2016bci}. A BCI framework based on recognition of imagined speech can provide a user with a more natural way for interaction with electronic devices \cite{hashim2017word}. Imagined speech Recognition here may be defined as the automated recognition of a given object, word or a letter from brain signals of the user. There are various techniques to measure brain signals ranging from invasive methods such as surgically implanted electrodes to the non-invasive EEG. In this work, we focus on non-invasive brain signals acquired using EEG technique. Despite of the low spatial resolution of EEG, it has been widely popular owing to it's simplistic nature and little to no discomfort to the user \cite{alsaleh2018discriminating}. 

The recognition of imagined speech using EEG has been attempted at various levels such as word, syllable, vowel imagination \cite{alsaleh2018discriminating, saha2019deep, nguyen2017inferring}. In their work, Gonzalez-Castaneda et al. \cite{gonzalez2017sonification} proposed an imagined word classification system using EEG comprising of $5$ words. They first applied a Common Average Reference method to improve the Signal to Noise ratio and then extracted features using the Discrete Wavelet Transform.  In \cite{hashim2017word}, classified $2$ words by using Mel Frequency Cepstral Coefficients and a k-Nearest Neighbor classifier. The majority of current envisioned speech recognition systems are heavily reliant on extracting meaningful handcrafted features, which requires in-depth domain expertise \cite{zhang2018cascade}. With advancements in the field of deep learning, researchers switched attention towards automatic extraction of features from raw EEG signals. The features extracted from the signals are such that they contain discriminatory information about the classification task. Zhang et al. \cite{zhang2018cascade} developed a combination of cascade and parallel Convolutional Recurrent Neural Network (CRNN) architecture for intention recognition using EEG signals. The EEG signals were processed by using 2D-CNN by constructing a 2-dimensional mesh as per the spatial information utilizing the electrode distribution. The features thereby extracted were fed to a LSTM for classification. For a $5$-class classification task, the authors reported an improvement of around $18\%$ over features extracted from frequency information.

In this work, we study the contribution of different EEG frequency bands ($\delta, \theta, \alpha, \beta, \gamma$) and signals acquired from different lobes of the brain (frontal, temporal, occipital and parietal) towards imagined speech recognition using EEG signals. The signals are first bandpass filtered according to different frequency bands as (a) $\delta : <4 Hz$, (b) $\theta : 4-7 Hz$, (c) $\alpha : 8-15 Hz$, (d) $\beta : 16-31 Hz$, and (e) $\gamma : >32 Hz$. These bandpass filtered signals are fed to a 1DCNN-LSTM based architecture. The purpose of the 1D-CNN block is to extract spatial features from the signals while the LSTM block is used in order to extract the temporal information. We also select specific lobes of the brain and signals from only the selected lobes are fed to the classifier and the classification performance is obtained. Based on our observation that the frequency bands $\theta$ and $\alpha$ have least contribution towards the classification of envisioned speech, we filter out these bands by using a bandreject filter to feed the the classifier. We show that such an approach improves the classification accuracy compared to using all the entire raw EEG signal. We demonstrate the efficacy of this approach on a publicly available envisioned speech dataset \cite{kumar2018envisioned} consisting of $3$ tasks - digit, character and images. We achieve classification accuracy of $85.93\%$, $87.27\%$ and $87.51\%$ for the three tasks respectively. 

\section{Dataset}

We use a publicly available envisioned speech dataset containing recordings from 23 participants aged between 15-40 years \cite{kumar2018envisioned}. EEG was recorded using Emotiv EPOC+ \cite{emotiv} wireless neuro headset. The headset consists of $14$ channels namely AF3, AF4, F3, F4, F7, F8, FC5, FC6, T7, T8, P7, P8, O1 and O2. The electrodes were placed at the scalp of the subject according to International 10-20 system of electrode placement as depicted in Figure \ref{fig:eegplacement}. In addition to the aforementioned electrodes, two reference electrodes CMS and DRL were placed above the ears. The signals were recorded at a sampling frequency of $2048$ Hz and then down-sampled to $128$ Hz. 

\begin{figure}[]
  \centering
  \centerline{\includegraphics[width=\linewidth]{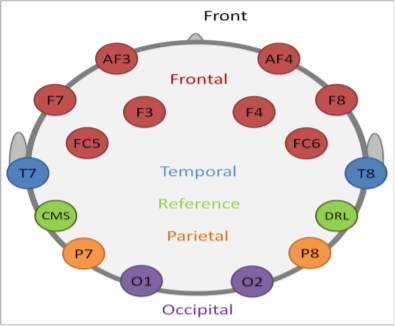}}
  \caption{Pictorial representation of the electrode placement for the data collection. The figure has been adapted from \cite{kumar2018envisioned}.} 
\label{fig:eegplacement}
\end{figure}

\begin{figure}[]
  \centering
  \centerline{\includegraphics[width=\linewidth]{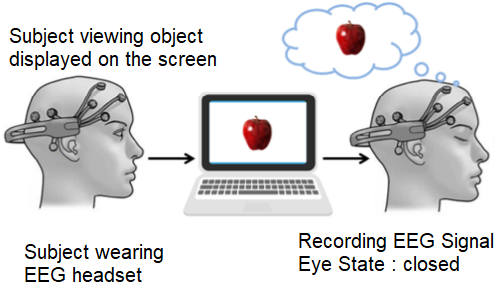}}
  \caption{Flow diagram of the envisioned speech recognition recording setup. The figure has been adapted from \cite{kumar2018envisioned}.} 
\label{fig:datacollection}
\end{figure}

As depicted in Figure \ref{fig:datacollection}, the subject is displayed an object on the screen. Then, subject is instructed to imagine the viewed object for $10$ seconds while keeping the eyes closed and in resting state. Before displaying the next image to the participant, a gap of $20$ seconds is maintained in order to ensure that the subject returns to the state of rest before displaying the next item. In this manner, three categories of prompts have been shown to the participants and the corresponding EEG has been recorded. 

The first category comprises of digits $0-9$, second comprises of $10$ uppercase English alphabets - A, C, F, H, J, M, P, S, T, Y. The last category consists of items used in daily life - Apple, Car, Dog, Gold, Mobile, Rose, Scooter, Tiger, Wallet and Watch. In total, $230 (23*10)$ EEG recordings per category (comprising of $10$ classes) have been collected.

\section{Experimental Setup}

\subsection{Methodology}

Classification was performed by taking the three subsets - digits, characters and images of the described dataset. The entire EEG signal recorded for a particular class is of $10$ seconds duration. As in \cite{kumar2021deep}, the EEG signals were divided into $250$ ms duration (i.e. 32 samples) with a sliding increment of $64$ ms (i.e. 8 samples). For analyzing each of the different frequency bands, the signals are filtered by using the following standards:
\begin{itemize}
    \item $\delta$ : low pass filter with a cutoff 4 Hz. 
    \item $\theta$ : bandpass filter between $4-7$ Hz.
    \item $\alpha$ : bandpass filter between $7-15$ Hz.
    \item $\beta$ : bandpass filter between $15-31$ Hz.
    \item $\gamma$ : high pass filter with a cutoff $31$ Hz.
\end{itemize}

The filtered signals are separately fed to the 1DCNN-LSTM model for classification. Out of the $14$ electrodes from which EEG signal was recorded,  we identify $8$ channels corresponding to the frontal lobe and $2$ channels corresponding to the temporal, parietal and occipital lobes. EEG signals corresponding to these different lobes are also separately used for classification in raw form as well as by obtaining different frequency band information.

\subsection{Model Architecture}
\begin{figure}[]
  \centering
  \centerline{\includegraphics[width=0.9\linewidth]{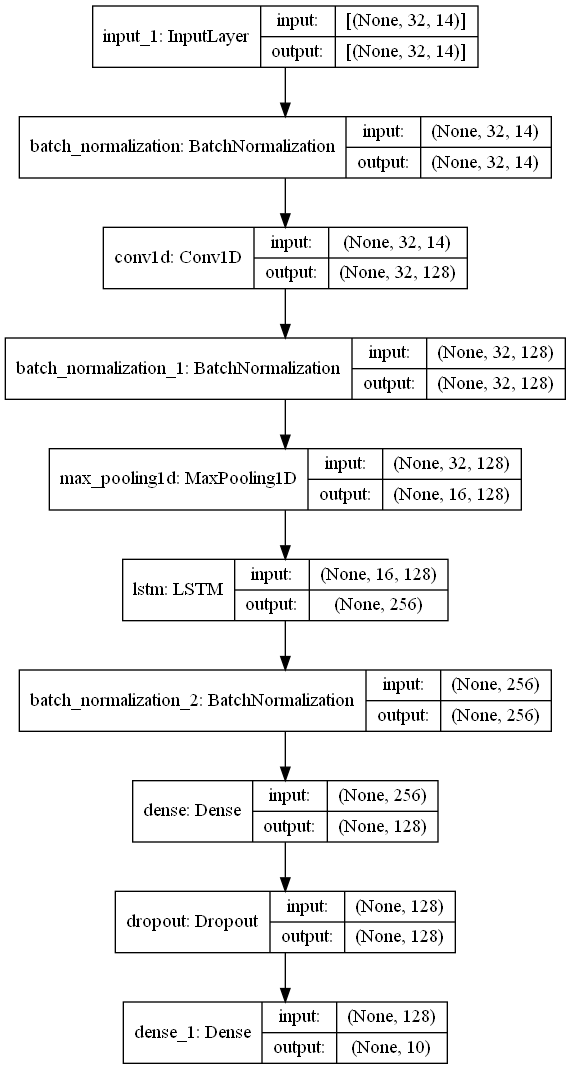}}
  \caption{Model Architecture for the envisioned speech recognition system.} 
\label{fig:model}
\end{figure}

The entire model architecture is depicted in Figure \ref{fig:model}. The input to the model is a matrix of dimension $32\times N_c$ where $32$ represents the timestamp and $N_c$ is the number of channels ($14$ when all channels are considered, $8$ if only frontal lobe channels are considered and $2$ if either of temporal, occipital or parietal lobe channels are considered). For the 1DCNN layer, $64$ filters of size $10$ with a stride of $1$ are taken in order to extract the spatial information from the input. Subsequently, maxpooling by $2$ is done. The spatial features extracted from the input are then fed to a LSTM block consisting of $256$ units to extract meaningful temporal information. This is then fed to a Dense layer comprising of $128$ neurons activated by the ReLU activation function and finally to the output layer comprising of $10$ neurons activated by the softmax activation function. 

\subsection{Experimental Details}
The data in each of the three tasks is split into a training set and a test set in a ratio of $80:20$. We employ a mini-batch training process by keeping a batch size of $128$. In order to prevent overfitting, an early stopping with a patience of $10$ while monitoring the validation accuracy is employed and a dropout of $50\%$ is used between the output layer and the penultimate layer of the CNN-LSTM architecture. Since the dataset is balanced among the different classes, categorical cross entropy loss is minimized using the Adam optimizer \cite{kingma2014adam}.

\section{Results}

Table \ref{tab:digit_results} lists the performance of EEG signals obtained from different lobes of the brain and bandpass filtered for different EEG frequency bands. In terms of different frequency bands, it can be seen that the low frequency $\delta$ band contains the most information followed by the $\gamma$ and $\beta$ bands. The performance of $\theta$ and $\alpha$ bands are significantly low compared to the other frequency bands. Thus, when the signal is bandpass filtered and the $\theta$ and $\alpha$ frequency bands are removed, we see a boost in classification performance. This is very well in line with the literature of different frequency bands and their normal occurrence with respect to the task at hand. Furthermore, it can also be seen that the frontal lobe contains most information and therefore the accuracy of signals when only the frontal lobe is considered is significantly high. In Tables \ref{tab:char_results} and \ref{tab:image_results}, the results of experiments performed on the character and image tasks of the dataset are presented. The trend observed in these two tasks are also similar to that observed for the digit classification task.

\begin{table}[]
\caption{Recognition accuracies for digit task}
\centering
\scalebox{0.8}{
\begin{tabular}{|c|c|c|c|c|c|c|c|}
\hline
\textbf{Part/Band} & $\delta$ & $\theta$ & $\alpha$ & $\beta$      & $\gamma$      & $\delta + \beta + \gamma$                      & \textbf{All}    \\ \hline\hline
\textbf{Frontal}   & 0.6934     & 0.3283     & 0.2651     & 0.5823          & 0.6318          & \textbf{0.7402}                         & 0.7244          \\ \hline
\textbf{Temporal}  & 0.2018     & 0.1149     & 0.1178     & 0.2824          & 0.3834          & 0.4692                                  & \textbf{0.4983} \\ \hline
\textbf{Occipital} & 0.1248     & 0.1021     & 0.1028     & \textbf{0.2749} & 0.2207          & 0.1266                                  & 0.2211          \\ \hline
\textbf{Parietal}  & 0.1426     & 0.1055     & 0.1124     & \textbf{0.3496} & \textbf{0.3435} & 0.2917                                  & 0.2993          \\ \hline
\textbf{All}       & 0.8433     & 0.5009     & 0.5354     & 0.6769          & 0.7478          & \cellcolor[HTML]{00B0F0}\textbf{0.8593} & \textbf{0.8582} \\ \hline
\end{tabular}

}
\label{tab:digit_results}
\end{table}

\begin{table}[]
\caption{Recognition accuracies for character task}
\centering
\scalebox{0.8}{
\begin{tabular}{|c|c|c|c|c|c|c|c|}
\hline
\textbf{Part/Band} & $\delta$ & $\theta$ & $\alpha$ & $\beta$      & $\gamma$      & $\delta + \beta + \gamma$                      & \textbf{All}    \\ \hline\hline
\textbf{Frontal}   & 0.6978     & 0.3441     & 0.3899     & 0.5678     & 0.6433          & \textbf{0.7871}                & 0.7521                                  \\ \hline
\textbf{Temporal}  & 0.1556     & 0.1034     & 0.1077     & 0.3154     & 0.3928          & \textbf{0.5221}                & \textbf{0.5221}                         \\ \hline
\textbf{Occipital} & 0.1391     & 0.1084     & 0.1261     & 0.3024     & \textbf{0.3456} & 0.1861                         & 0.2522                                  \\ \hline
\textbf{Parietal}  & 0.1381     & 0.1089     & 0.1136     & 0.3191     & \textbf{0.3271} & 0.2445                         & 0.2893                                  \\ \hline
\textbf{All}       & 0.8511     & 0.5405     & 0.1311     & 0.6872     & 0.7197          & 0.8727 & \cellcolor[HTML]{00B0F0}\textbf{0.8735} \\ \hline
\end{tabular}

}
\label{tab:char_results}
\end{table}

\begin{table}[]
\caption{Recognition accuracies for image task}
\centering
\scalebox{0.8}{
\begin{tabular}{|c|c|c|c|c|c|c|c|}
\hline
\textbf{Part/Band} & $\delta$ & $\theta$ & $\alpha$ & $\beta$      & $\gamma$      & $\delta + \beta + \gamma$                      & \textbf{All}    \\ \hline\hline
\textbf{Frontal}   & 0.7155     & 0.2021     & 0.3756     & 0.5634          & 0.6236          & \textbf{0.8042}                         & 0.7593                         \\ \hline
\textbf{Temporal}  & 0.1903     & 0.1021     & 0.1296     & 0.3144          & 0.3731          & \textbf{0.5153}                         & 0.4965                         \\ \hline
\textbf{Occipital} & 0.1323     & 0.1156     & 0.1095     & 0.1906          & \textbf{0.3503} & 0.2079                                  & 0.2157                         \\ \hline
\textbf{Parietal}  & 0.1469     & 0.0962     & 0.2006     & \textbf{0.3361} & 0.3144          & \textbf{0.3364}                         & 0.3231                         \\ \hline
\textbf{All}       & 0.8396     & 0.4961     & 0.5908     & 0.6575          & 0.7262          & \cellcolor[HTML]{00B0F0}\textbf{0.8751} & 0.8699 \\ \hline
\end{tabular}
}
\label{tab:image_results}
\end{table}

\begin{figure}[]
  \centering
  \centerline{\includegraphics[width=\linewidth]{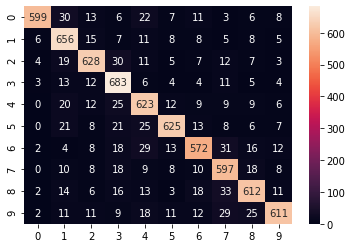}}
  \caption{Confusion matrix for the digit task.} 
\label{fig:eegplacement}
\end{figure}

\begin{figure}[]
  \centering
  \centerline{\includegraphics[width=\linewidth]{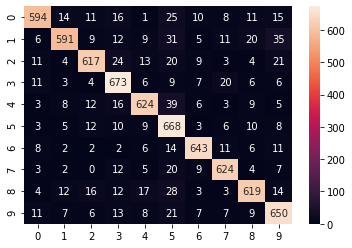}}
  \caption{Confusion matrix for the character task. 0-9 represents in order the characters A, C, F, H, J, M, P, S, T, Y.} 
\label{fig:eegplacement}
\end{figure}

\begin{figure}[]
  \centering
  \centerline{\includegraphics[width=\linewidth]{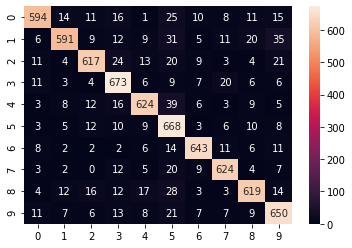}}
  \caption{Confusion matrix for the image task. 0-9 represents in order the images of Apple, Car, Dog, Gold, Mobile, Rose, Scooter, Tiger, Wallet and Watch.} 
\label{fig:eegplacement}
\end{figure}

\begin{table}[]
\caption{Comparison with previous work. (The results of other approaches have been adapted from \cite{kumar2021deep}}
\centering
\scalebox{0.8}{
\begin{tabular}{|c|c|c|c|c|}
\hline
\textbf{Author} & \textbf{Approach} & \textbf{Character} & \textbf{Digit} & \textbf{Image}    \\ \hline\hline
Tirupattur et.al. \cite{tirupattur2018thoughtviz} & CNN & $71.2\%$ & $72.9\%$ & $73.0\%$ \\ \hline

Jolly et.al. \cite{jolly2019universal} & CNN,GRU & $-$ & $-$ & $77.4\%$ \\ \hline

Kumar et.al. \cite{kumar2018envisioned} & Random Forest & $66.9\%$ & $68.5\%$ & $65.7\%$ \\ \hline

For Comparison \cite{kumar2021deep} & CNN alone & $71.0\%$ & $66.4\%$ & $72.0\%$ \\ \hline

For Comparison \cite{kumar2021deep} & Stacked LSTM & $84.2\%$ & $75.7\%$ & $82.4\%$ \\ \hline

For Comparison \cite{kumar2021deep} & CNN+LSTM & $87.1\%$ & $82.8\%$ & $86.6\%$ \\ \hline

Kumar et.al. \cite{kumar2021deep} & CNN+LSTM+MV & \textbf{90.1\%} & $85.1\%$ & \textbf{89.4\%} \\ \hline

Proposed & ($\delta + \beta + \gamma$) + CNN$-$LSTM & $87.3\%$ & \textbf{85.9\%} & $87.5\%$ \\ \hline

\end{tabular}
}
\label{tab:comparison}
\end{table}

\section{Discussion \& Conclusion}

Due to the rapid advancement in low-cost EEG measurement systems, there has been a proportionate increase in real-world Brain Computer Interface applications. The recognition of imagined speech is an important task for people suffering from voice disorders or those with conditions such as locked-in syndrome. Moreover, it is also of importance for people with high privacy needs such as in public places where speech is not an ideal medium of communication. In spite of the tremendous possible use cases, there has been scarce work in this field compared to other motor tasks such as grasp and lift, movement etc. Since speech is a natural way of communication among people and also for interaction with other electronic devices, it is expected that a system based on decoding imagined speech should be of extreme importance. Most work surrounding the recognition of imagined speech has been done by using traditional signal processing based approaches such as extracting features like common spatial patterns (CSP), auto-regression coefficients, low-level descriptors such as energy, entropy etc. along with machine learning classifiers such as SVM, Random Forest k-NN, and Logistic Regression. In more recent works, there has been some focus on leveraging the advancement in Deep Learning by using architectures such as CNN, LSTM, GRU etc. for learning the complexities in the EEG signals. 

In this work, we aim to explore the role of signals from different lobes of the brain (frontal, occipital, parietal and temporal) along with different EEG frequency bands ($\delta, \theta, \alpha, \beta, \gamma$) for imagined speech recognition. We use parallel bandpass filters to filter the EEG signal into corresponding frequency bands and feed it to a 1DCNN-LSTM based deep learning architecture for imagined speech and object recognition. The 1DCNN block extracts spatial features from the raw EEG signals and the LSTM block models the temporal information over the extracted features.  We observe that the frequency bands $\theta$ and $\alpha$ do not significantly contribute towards imagined speech recognition and therefore, we suggest bandreject filter for preprocessing the EEG signals before using them for the classification of imagined speech. Such an approach has been found to boost the classification performance on comparison with feeding directly the raw EEG signals to the 1DCNN-LSTM architecture. Furthermore, we also observe that the frontal lobe is the most significant contributor compared to other lobes of the brain. Both the observations are well in line with the literature on the functions of the frequency bands and the lobes. We perform experiments on a publicly available imagined speech dataset and achieve classification accuracies of $85.93\%$, $87.27\%$ and $87.51\%$ for the task of classifying digits, characters and images respectively.

\bibliographystyle{IEEEtran}
\bibliography{refs.bib}

\end{document}